\documentclass[journal]{IEEEtran}
\ifCLASSINFOpdf
   \usepackage[pdftex]{graphicx}
\else
\fi
\usepackage{amsmath}
\usepackage{graphicx}
\usepackage{array}
\usepackage{verbatim}
\usepackage{subfigure}
\usepackage[utf8]{inputenc}
\usepackage[table]{xcolor}
\usepackage{float}
\usepackage{url}
\usepackage{multirow}
\usepackage{tcolorbox}
\usepackage{hyperref}

\usepackage{comment}
\usepackage[utf8]{inputenc}
\inputencoding{utf8}
\begin{document}

%
% paper title
% Titles are generally capitalized except for words such as a, an, and, as,
% at, but, by, for, in, nor, of, on, or, the, to and up, which are usually
% not capitalized unless they are the first or last word of the title.
% Linebreaks \\ can be used within to get better formatting as desired.
% Do not put math or special symbols in the title.
\title{MEG Source Localization via Deep Learning}

% author names and affiliations
% transmag papers use the long conference author name format.

\author{Dimitrios Pantazis and Amir Adler}

%\author{Amir Adler\thanks{A. Adler and T. Poggio are with the Center for Brains, Minds and Machines (CBMM), MIT, Cambridge, MA 02139, USA, Email: adleram@mit.edu}, Mauricio Araya-Polo\thanks{M. Araya-Polo is with Shell International Exploration and Production Inc. (now with Total EP RT USA), Houston, TX 77082, USA. Email: mauricio.araya@shell.com}, Tomaso Poggio}

%\thanks{Manuscript received December 1, 2012; revised August 26, 2015. 
%Corresponding author: M. Shell (email: http://www.michaelshell.org/contact.html).}

% The paper headers
\markboth{}%
{Shell \MakeLowercase{\textit{et al.}}: Bare Demo of IEEEtran.cls for IEEE Transactions on Magnetics Journals}
% The only time the second header will appear is for the odd numbered pages
% after the title page when using the twoside option.
% 
% *** Note that you probably will NOT want to include the author's ***
% *** name in the headers of peer review papers.                   ***
% You can use \ifCLASSOPTIONpeerreview for conditional compilation here if
% you desire.

\maketitle
%\IEEEtitleabstractindextext{%

%\input{abstract.tex}

% Note that keywords are not normally used for peerreview papers.
%\begin{IEEEkeywords}
%Image reconstruction, Tomography, Spectrogram, Convolutional Neural Network, %Super-resolution
%\end{IEEEkeywords}}

% make the title area

% To allow for easy dual compilation without having to reenter the
% abstract/keywords data, the \IEEEtitleabstractindextext text will
% not be used in maketitle, but will appear (i.e., to be "transported")
% here as \IEEEdisplaynontitleabstractindextext when the compsoc 
% or transmag modes are not selected <OR> if conference mode is selected 
% - because all conference papers position the abstract like regular
% papers do.
\IEEEdisplaynontitleabstractindextext
% \IEEEdisplaynontitleabstractindextext has no effect when using
% compsoc or transmag under a non-conference mode.

% For peer review papers, you can put extra information on the cover
% page as needed:
% \ifCLASSOPTIONpeerreview
% \begin{center} \bfseries EDICS Category: 3-BBND \end{center}
% \fi
%
% For peerreview papers, this IEEEtran command inserts a page break and
% creates the second title. It will be ignored for other modes.
\IEEEpeerreviewmaketitle

\begin{abstract}
We present a deep learning solution to the problem of localization of magnetoencephalography (MEG) brain signals. The proposed deep model architectures are tuned for single and multiple time point MEG data, and can estimate varying numbers of dipole sources. Results from simulated MEG data on the cortical surface of a real human subject demonstrated improvements against the popular RAP-MUSIC localization algorithm in specific scenarios with varying SNR levels, inter-source correlation values, and number of sources. Importantly, the deep learning models had robust performance to forward model errors and a significant reduction in  computation time, to a fraction of 1 ms, paving the way to real-time MEG source localization. 
\end{abstract}

\section{Introduction}
\IEEEPARstart{A}{\lowercase{ccurate}} localization of functional brain activity holds promise to enable novel treatments and assistive technologies that are in critical need by our aging society. The ageing of the world population has increased the prevalence of age-related health problems, such as physical injuries, mental disorders, and stroke, leading to severe consequences for patients, families, and the health care system. Emerging technologies can improve the quality of life of patients by i) providing effective neurorehabilitation, and ii) enabling independence in everyday tasks. The first challenge may be addressed by designing neuromodulatory interfacing systems that can enhance specific cognitive functions or treat specific psychiatric/neurological pathologies. Such systems could be driven by real-time brain activity to selectively modulate specific neurodynamics using approaches such as transcranial magnetic stimulation \cite{VALEROCABRE2017381,KOHL2019355} or focused ultrasound \cite{FOLLONI20191109,Darrow2019}. The second challenge may be addressed by designing effective brain-machine interfaces (BMI). Common BMI control signals rely on primary sensory- or motor-related activation. However, these signals only reflect a limited range of cognitive processes. Higher-order cognitive signals, and specifically those from prefrontal cortex that encode goal-oriented tasks, could lead to more robust and intuitive BMI \cite{MIN2017585,PICHIORRI2020101}. 

Both neurorehabilitation and BMI approaches necessitate an effective and accurate way of measuring and localizing functional brain activity in real time. This can be achieved by electroencephalography (EEG) \cite{ilmoniemi2019brain,niedermeyer_niedermeyers_2012} and MEG \cite{hamalainen_magnetoencephalographytheory_1993, baillet_magnetoencephalography_2017,DARVAS2004S289}, two non-invasive electrophysiological techniques. EEG uses an array of electrodes placed on the scalp to record voltage fluctuations, whereas MEG uses sensitive magnetic detectors called superconducting quantum interference devices (SQUIDs) \cite{kleiner_superconducting_2004} to measure the same primary electrical currents that generate the electric potential distributions recorded in EEG. Since EEG and MEG capture the electromagnetic fields produced by neuronal currents, they provide a fast and direct index of neuronal activity. However, existing MEG/EEG source localization methods offer limited spatial resolution, confounding the origin of signals that could be used for neurorehabilitation or BMI, or are too slow to compute in real time. 

Deep learning (DL) \cite{Goodfellow-et-al-2016} offers a promising new approach to significantly improve source localization in real time. A growing number of works successfully employ DL to achieve state-of-the-art image quality for inverse imaging problems, such as X-ray computed tomography (CT) \cite{gupta_cnn-based_2018, jin_deep_2017,flohr_deep_2017}, magnetic resonance imaging (MRI) \cite{8962949,wang_accelerating_2016,schlemper_deep_2018}, positron emission tomography (PET) \cite{gong_pet_2018,kim_penalized_2018}, image super-resolution \cite{dong_image_2016,kim_accurate_2016,lim_enhanced_2017}, photoacoustic tomography \cite{hauptmann_model-based_2018}, synthetic aperture radar (SAR) image reconstruction  \cite{yonel_deep_2018,budillon_sar_2019} and seismic tomography \cite{Araya2018}. Here, we develop and present a novel DL solution to localize neural sources, and assess its accuracy and robustness with simulated MEG data on the sensor geometry of the whole-head Elekta Triux MEG system and source space the cortical surface extracted from a structural MRI scan of a real human subject. While we focus on MEG, the same approaches are directly extendable to EEG, enabling a portable and affordable solution to source localization. The conceptual novelties of the proposed DL methods are: i) state-of-the-art localization accuracy of multiple sources from noisy MEG measurements, and ii) very fast computation time (fraction of a millisecond) for inference of the source locations. 
\begin{figure*}[ht]
\centering
\includegraphics[width=0.85\textwidth]{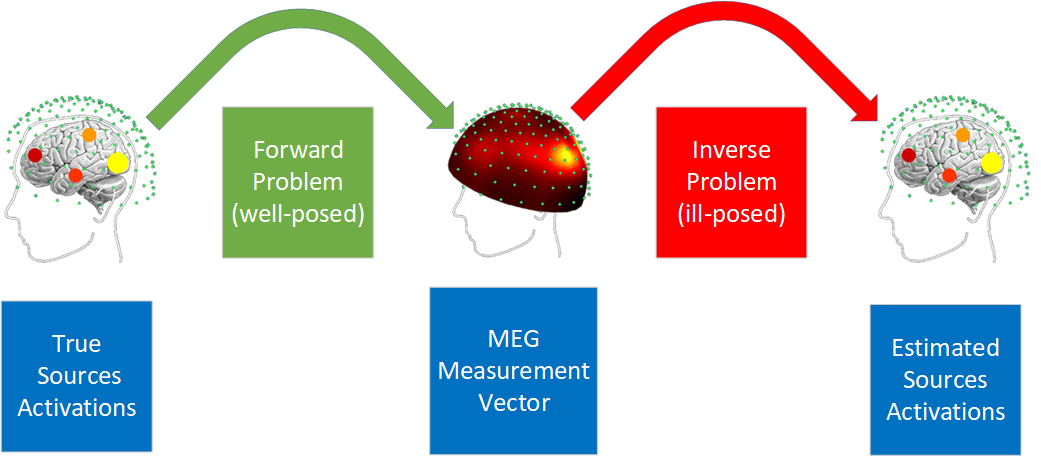}
\caption{MEG forward and inverse problems: in the forward problem a well-posed model maps the true sources activation to the MEG measurement vector. In the inverse (and ill-posed) problem, an inverse operator maps the measurement vector to the estimated sources activation.}
\label{fig:forward_inverse}
\end{figure*}

\section{Background on MEG Source Localization}
Two noninvasive techniques, MEG and EEG, measure the electromagnetic signals emitted by the human brain and can provide a fast and direct index of neural activity suitable for real-time applications. The primary source of these electromagnetic signals is widely believed to be the electrical currents flowing through the apical dendrites of pyramidal neurons in the cerebral cortex. Clusters of thousands of synchronously activated pyramidal cortical neurons can be modeled as an equivalent current dipole (ECD). The current dipole is therefore the basic element used to represent neural activation in MEG and EEG localization methods. 

In this section we briefly review the notations used to describe measurement data, forward matrix, and sources, and formulate the problem of estimating current dipoles. Consider an array of $M$ MEG or EEG sensors that measures data from a finite number $Q$ of equivalent current dipole (ECD) sources emitting signals $\{s_q(t)\}^{Q}_{q=1}$ at locations $\{\mathbf p_q\}^{Q}_{q=1}$. Under these assumptions, the $M\times 1$ vector of the received signals by the array is given by:
\begin{equation}
%\label{basic_moldel}
\mathbf y(t) = \sum_{q=1}^{Q} \mathbf l(\mathbf p_q)  s_{q}(t) +\mathbf n(t),
\label{eq:snapshot1}
\end{equation}
where  $\mathbf l(\mathbf p_q)$ is the topography of the dipole at location $\mathbf p_q$  and $\mathbf n(t)$  is the additive noise. The topography $\mathbf l(\mathbf p_q)$, is given by: 
\begin{equation}
\label{topography_model}
\mathbf l(\mathbf p_q) = \mathbf L(\mathbf p_q)  \mathbf{q},
\end{equation}
where $\mathbf L (\mathbf p_q)$ is the $M\times 3$ forward matrix at location $\mathbf p_q$ and $\mathbf q$ is the $3\times 1$ vector of the orientation of the ECD source. Depending on the problem, the orientation $\mathbf q$ may be  known, referred to as \textit{fixed-oriented} dipole, or it may be  unknown, referred to as \textit{ freely-oriented} dipole.

Assuming that the array is sampled  $N$ times at $t_1,...,t_N$, the matrix $\mathbf Y$ of the sampled signals can be expressed as: 
\begin{equation}
\label{basic_equation}
\mathbf Y =\mathbf A(\mathbf P) \mathbf S +\mathbf N, 
\end{equation} 
where  $\mathbf Y$ is the $M\times N$ matrix of the received signals:
\begin{equation}
\mathbf Y= [\mathbf y(t_1), ..., \mathbf y (t_N)],
\label{eq:snapshot4}
\end{equation}
$\mathbf A(\mathbf P)$ is the $M\times Q$ mixing matrix of the topography vectors at the $Q$ locations $\mathbf P=[\mathbf p_1,...,\mathbf p_Q]$:
\begin{equation}
\mathbf A(\mathbf P)=[\mathbf l(\mathbf p_1), ..., \mathbf l(\mathbf p_Q)],
\label{eq:snapshot5}
\end{equation}
$\mathbf S$ is the $Q\times N$ matrix of the sources:
\begin{equation}
\mathbf S= [\mathbf s(t_1), ..., \mathbf s (t_N)],
\label{eq:snapshot6}
\end{equation}
with $\mathbf s(t)=[s_1(t),..., s_Q(t)]^{T}$, and $\mathbf N$ is the $M\times N$ matrix of noise:
\begin{equation}
\mathbf N =[\mathbf n(t_1),..., \mathbf n(t_N)].
\label{eq:snapshot8}
\end{equation}

Mathematically, the localization problem can be cast as an optimization problem of computing the location and moment parameters of the set of dipoles whose field best matches the MEG/EEG measurements in a least-squares (LS) sense \cite{mosher_multiple_1992}. In this paper we focus on solutions that solve for a small parsimonious set of dipoles and avoid the ill-posedness associated with imaging methods that yield distributed solutions, such as minimum-norm \cite{hamalainen_magnetoencephalographytheory_1993}. Solutions that estimate a small set of dipoles include the \textit{dipole fitting} and \textit{scanning} methods. Dipole fitting methods solve the optimization problem directly using techniques that include gradient descent, Nedler-Meade simplex algorithm, multistart, genetic algorithm, and simulated annealing \cite{huang_multi-start_1998,uutela_global_1998,khosla_spatio-temporal_1997,jiang_comparative_2003}. However, these techniques remain unpopular because they converge to a suboptimal local optimum or are too computationally expensive.

An alternative approach is scanning methods, which use adaptive spatial filters to search for optimal dipole positions throughout a discrete grid representing the source space \cite{darvas_mapping_2004}. Source locations are then determined as those for which a metric (localizer) exceeds a given threshold. While these approaches do not lead to true least squares solutions, they can be used to initialize a local least squares search. The most common scanning methods are beamformers \cite{LCMV,VerbaRobinson} and MUSIC \cite{mosher_multiple_1992}, both widely used for bioelectromagnetic source localization, but they assume uncorrelated sources. When correlations are significant, they result in partial or complete cancellation of correlated (also referred to as synchronous) sources) and distort the estimated time courses. Several multi-source extensions have been proposed for synchronous sources \cite{CohBF, CorrBF, NullBF, DCBF, EDCBF, MultiLCMV, MultiBeamformers, POP-MUSIC, WedgeMUSIC}, however they require some a-priori information on the location of the synchronous sources, are limited to the localization of pairs of synchronous sources, or are limited in their performance.\\
\indent One important division of the scanning methods is whether they are \textit{non-recursive} or \textit{recursive}. The original Beamformer \cite{LCMV,VerbaRobinson} and MUSIC \cite{mosher_multiple_1992} methods are non-recursive and require the identification of the largest local maxima in the localizer function to find multiple dipoles. Some multi-source variants are also non-recursive (e.g. \cite{hui_identifying_2010, CorrBF, DCBF, EDCBF}), and as a result they use brute-force optimization, assume that the approximate locations of the neuronal sources have been identified a priori, or still require the identification of the largest local maxima in the localizer function. To overcome these limitations, non-recursive methods have recursive counterparts, such as RAP MUSIC \cite{Mosher1999}, TRAP MUSIC \cite{makela_truncated_2018}, Recursive Double-Scanning MUSIC \cite{makela_locating_2017}, and RAP Beamformer \cite{ilmoniemi2019brain}. The idea behind the recursive execution is that one finds the source locations iteratively at each step, projecting out the topographies of the previously found dipoles before forming the localizer for the current step \cite{Mosher1999,ilmoniemi2019brain}. In this way, one replaces the task of finding several local maxima with the easier task of finding the global maximum of the localizer at each iteration step. While recursive methods generally perform better than their non-recursive counterparts, they still suffer from several limitations, including limited performance, the need for high signal-to-noise ratio (SNR), non-linear optimization of source orientation angles and source amplitudes, or inaccurate estimation as correlation values increase. The are also computationally expensive due to the recursive estimation of sources.

\section{Background on Deep Learning}
\begin{figure*}
\centering
\includegraphics[width=\textwidth]{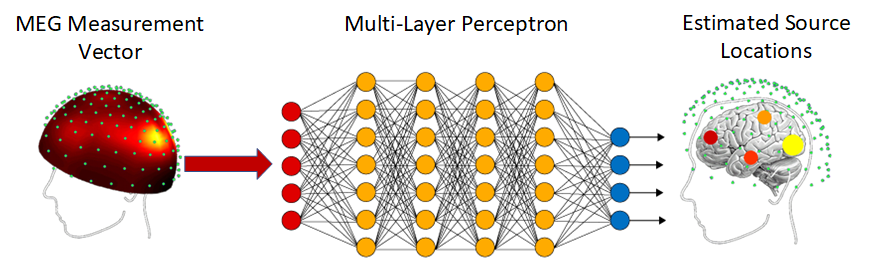}
\caption{Illustration of the MLP-based MEG source localization solution: the end-to-end inversion operator performs mapping from the MEG measurement space to the source locations space.}
\label{fig:DNN}
\end{figure*}
Inverse problems in signal and image processing were traditionally solved using analytical methods, however, recent DL~\cite{Goodfellow-et-al-2016} solutions, as exemplified in  \cite{8253590,Deep_MRI_8962949}, provide state-of-the-art results for numerous problems including x-ray computed tomography, magnetic resonance image reconstruction, natural image restoration (denoising, super-resolution, debluring), synthetic aperture radar image reconstruction and hyper-spectral unmixing, among others. In the following we review DL principles, which form the basis for the DL-based solutions to MEG source localization presented in the next section, including concepts such as network layers and activation functions, empirical risk minimization, gradient-based learning, and regularization.\\
\indent DL is a powerful class of data-driven machine learning algorithms for supervised, unsupervised, reinforcement and generative tasks. DL algorithms are built using Deep Neural Networks (DNNs), which are formed by a hierarchical composition of non-linear functions (layers). {The main reason for the success of DL is the ability to train very high capacity (i.e hypothesis space) networks using very large datasets, often leading to robust \textit{representation learning} \cite{6472238} and good \textit{generalization} capabilities in numerous problem domains}. Generalization is defined as the ability of an algorithm to perform well on unseen examples. In statistical learning terms an algorithm $\mathcal{A:X\rightarrow Y}$ is learned using a training dataset $\mathcal{S} = \{(x_1, y_1), . . . ,(x_N, y_N)\}$ of size $N$, where $\textbf{x}_i\in\mathcal{X}$ is a data sample and $y_i\in\mathcal{Y}$ is the corresponding label (for example, source location coordinates). Let $\mathcal{P(X,Y)}$ be the true distribution of the data, then the expected risk is defined by:
\begin{equation}
    \mathcal{R(A)}=E_{x,y \sim \mathcal{P(X,Y)}}[\mathcal{L(A}(x),y)],
\end{equation} 
where $\mathcal{L}$ is a loss function that measures the misfit between the algorithm output and the data label. The goal of DL is to find an algorithm $\mathcal{A}$ within a given capacity (i.e. function space) that minimizes the expected risk, however, the expected risk cannot be computed since the true distribution is unavailable. Therefore, the empirical risk is minimized instead: 
\begin{equation}
\mathcal{R_E(A)}=\frac{1}{N}\sum_{i=1}^{N}\mathcal{L(A}(x_i),y_i),    
\end{equation}
 which approximates the statistical expectation with an empirical mean computed using the training dataset. The \textit{generalization gap} is defined as the difference between the expected risk to the empirical risk: $\mathcal{R(A)} - \mathcal{R_E(A)}$. By using large training datasets and high capacity algorithms, DL has been shown to achieve a low generalization gap, where an approximation of the expected risk is computed using the learned algorithm and a held-out testing dataset $\mathcal{T} = \{(x_1, y_1), . . . ,(x_M, y_M)\}$ of size $M$, such that $\mathcal{S \cap T} = \emptyset.$

In the following subsections we describe the main building blocks of DNNs, including  multi-layer perceptron and convolutional neural networks. 
\subsection{Multi-Layer Perceptron (MLP)}
The elementary building block of the MLP is the \textit{Perceptron}, which computes a non-linear scalar function, termed \textit{activation}, of an input  $\mathbf{x}\in\mathbf{R}^n$, as follows:

\begin{equation}
y = f(\mathbf{w}^T\mathbf{x}+b),  
\end{equation}
where $\mathbf{w}$ is a vector of weights  and $b$ is a scalar bias. A common activation function is the Rectified Linear Unit (ReLU) \cite{Goodfellow-et-al-2016}, defined as:
\begin{equation*}
f(z)=
\begin{cases}
  z & \text{for }z>0\\    
  0 & \text{for }z\le0    
\end{cases},
\end{equation*}
and in this case the perceptron is given by:
\begin{equation*}
y=
\begin{cases}
  \mathbf{w}^T\mathbf{x}+b & \text{for }\mathbf{w}^T\mathbf{x}+b >0\\    
  0 & \text{for }\mathbf{w}^T\mathbf{x}+b \le0    
\end{cases},
\end{equation*}
other common activation functions are \textit{LeakyReLU},  \textit{sigmoid} and the \textit{tanh} (i.e. hyperbolic tangent). A single layer of perceptrons is composed of multiple perceptrons, all connected to the same input vector $\mathbf{x}$, with a unique weight vector and bias, per perceptron. A single layer of perceptrons can be formulated in matrix form, as follows:
\begin{equation}
\textbf{y} = f(\textbf{Wx+b}),  
\end{equation}
where each row of the matrix $\textbf{W}$ corresponds to the weights of one perceptron, and each element of the vector $\textbf{b}$ corresponds to the bias of one perceptron. The MLP is composed of multiple layers of perceptrons, such that the output of each layer becomes the input to the next layer. Such hierarchical composition of $k$ non-linear functions is formulated as follows:
\begin{equation}
\mathbf{F}(\mathbf{x};\Theta)=f_k(f_{k-1}(\cdots f_2(f_1(\mathbf{x};\theta_1);\theta_2);\theta_{k-1});\theta_k),
\label{eq:MLP_Function}
\end{equation}
where $\theta_i=[\textbf{W}_i,\textbf{b}_i]$ are the parameters (i.e weights and biases) of the \textit{i}-th layer and $\Theta=[\theta_1,\theta_2,\dots,\theta_k]$ is the set of all network parameters.
\\
\indent In the supervised learning framework, the parameters $\Theta$ are learned by minimizing the empirical risk, computed over the training dataset $\mathcal{S}$. The empirical risk can be regularized in order to improve DNN generalization, by mitigating over-fitting of the learned parameters to the training data. The regularized empirical risk is given by
\begin{equation}
\mathbf{J}(\Theta)=\frac{1}{N}\sum_{i=1}^{N}\mathcal{L}(\mathbf{F}(\mathbf{x}_i;\Theta),y_i)+ \alpha \mathbf{R}(\Theta), 
\label{eq:regularized_emp_risk}
\end{equation}
\noindent where $\alpha\ge 0$ controls the weight of the regularization term, which is often chosen as Tikhonov regularization $\mathbf{R}(\Theta)=\|\Theta\|^2_2$ or $L_1$ regularization $\mathbf{R}(\Theta)=\|\Theta\|_1$ that promotes sparsity of the network parameters. The optimal set of parameters $\Theta^{*}$ are obtained by solving
\begin{equation}
\Theta^{*}=\arg \min_{\Theta}\mathbf{J}(\Theta),
\label{eq:opt_params}
\end{equation}
where the minimization of the empirical risk  is typically performed by iterative gradient-based algorithms, such as the stochastic gradient descent (SGD)
\begin{equation}
\hat{\Theta}_{t+1}=\hat{\Theta}_{t}-\lambda\nabla{_\Theta}\mathbf{J}(\Theta),
\label{eq:SGD}
\end{equation}
where $\hat{\Theta}_{t}$ is the estimate of $\Theta^{*}$ at the t-\textit{th} iteration, $\lambda>0$ is the learning rate, and the approximated gradient $\nabla{_\Theta}\mathbf{J}(\Theta)$ is computed by the \textit{back-propagation} algorithm using a small random subset of examples from the training set $\mathcal{S}$.

\subsection{Convolutional Neural Networks}
Convolutional Neural Networks (CNNs) were originally developed for processing input images, using the weight sharing principle of a convolutional kernel that is convolved with input data. The main motivation is to reduce significantly the number of required learnable parameters, as compared to processing a full image by perceptrons, namely, allocating one weight per pixel for each perceptron. A CNN is composed of one or more convolutional layers, where each layer is composed of one of more learnable kernels. For a 2D input $I(i,j)$, a convolutional layer performs the convolution\footnote{Some ML libraries implement the cross-correlation operation.} between the input to the kernel(s)
\begin{equation}
    C(i,j)=(K * I)(i,j)=\sum_{m,n}W(m,n)I(i-m,j-n),
\end{equation}
where $W(m,n)$ is the kernel and $C(i,j)$ is the convolution result. A bias $b$ is further added to each convolution results, and an activation function $f()$ is applied, to obtain the \textit{feature map} $F(i,j)$ given by
\begin{equation}
    F(i,j)=f(C(i,j)+b).
\end{equation}
A convolutional layer with $K$ kernels produces $K$ feature maps, where kernels of 1D, 2D or 3D are commonly used. Convolutional layers are often immediately followed by sub-sampling layers, such as \textit{MaxPooling} that decimates information by picking the maximum value within a given array of values, or \textit{AveragePooling} that replaces a given array of values by their mean. CNN networks are typically composed by a cascade of convolutional layers, optionally followed by fully-connected (FC) layers, depending on the required task. A frequently used architecture is the convolutional encoder-decoder, which learns a low-dimensional representation of the input (i.e. encoding), which is further utilized to reconstruct the output (i.e. decoding). Convolutional encoder-decoder architectures can be utilized for representation learning \cite{6472238} in an unsupervised learning framework (i.e. convolutional auto-encoder \cite{CAE}), or for complex output data reconstructions. A well-known convolutional encoder-decoder architecture is the \textit{U-Net} \cite{RFB15a}, originally proposed for medical image segmentation. Another high successful CNN  architecture, termed \textit{ResNet}, employs residual blocks with skip connections that enable the training of very deep networks \cite{He_2016_CVPR}.

\begin{figure*}
\centering
\includegraphics[width=\textwidth]{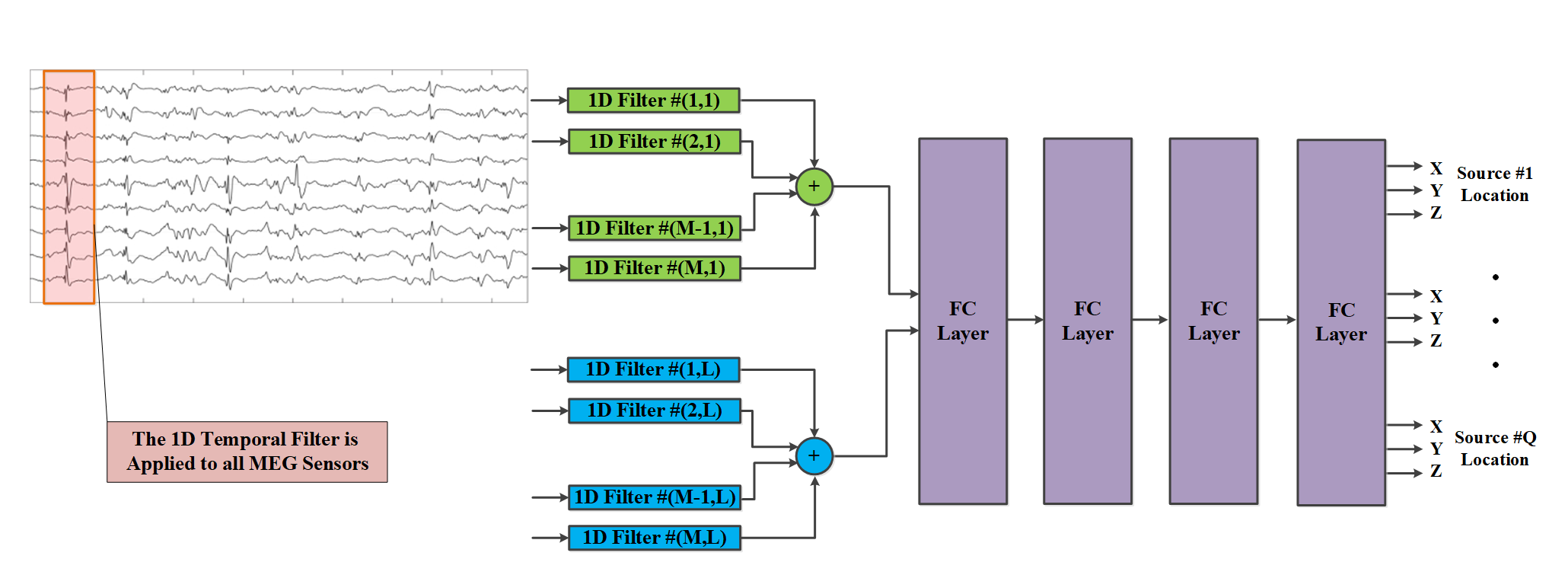}
\caption{Illustration of the CNN-based MEG source localization solution: the end-to-end inversion operator performs mapping from the MEG time-series measurement space, using a bank of $L$ space-time filters and four fully connected (FC) layers, to the source locations space.}
\label{fig:CNN}
\end{figure*}

\section{Deep Learning for MEG Source Localization}
In this section we present the proposed deep neural network (DNN) architectures and training data generation workflow.
\subsection{MLP for Single-Snapshot Source Localization}
MEG source localization is computed from sensor measurements using either a single snapshot (i.e. a single time sample) or multiple snapshots. The single snapshot case is highly challenging for popular MEG localization algorithms, such as MUSIC \cite{mosher_multiple_1992}, RAP-MUSIC \cite{Mosher1999}, and RAP-Beamformer \cite{ilmoniemi2019brain}, all of which rely on the data covariance matrix. A single snapshot estimation of the covariance matrix is often insufficient for good localization accuracy of multiple simultaneously active sources, especially in low and medium signal-to-noise ratios (SNR). Since the input in this case is a single measurement MEG vector, we implemented four layer MLP-based architectures, where the input FC layer maps the $M$-dimensional snapshot vector to a  higher dimensional vector and the output layer computes the source(s) coordinates in 3D, as illustrated in Figure \ref{fig:DNN}. We refer to this model as \textit{DeepMEG-MLP}. We implemented three DeepMEG-MLP models, corresponding to $Q=1,2,$ and $3$ sources, as summarized in Table \ref{tab:Architectures_Summary}.

\subsection{CNN for Multiple-Snapshot Source Localization}
For multiple consecutive MEG snapshots we implemented a CNN-based architecture with five layers, in which the first layer performs 1D convolutions on the input data and the resulting 1D feature maps are processed by three  FC layers with sigmoid activation, and an output FC layer which computes the source locations. We refer to this model as \textit{DeepMEG-CNN}, as illustrated in Figure \ref{fig:CNN}. The 1D convolutional layer forms a bank of $L=32$ space-time filters (which can also be interpreted as  beamformers \cite{7859320}). Each 1D temporal filter spans $T=5$ time samples. A different 1D filter is applied to the time course of each of the $M$ sensors with uniquely learned coefficients. We implemented three DeepMEG-CNN models, corresponding to $Q=1,2,$ and $3$ sources, as summarized in Table \ref{tab:Architectures_Summary}.

\begin{table*}[ht!]
\caption{Evaluated DeepMEG Architectures}
\centering
\begin{tabular}{|l|l|c|c|c|c|c|c|}
\hline
 Input & Network & $1^{st}$ Layer &  $2^{nd}$ Layer & $3^{rd}$ Layer &$4^{th}$ Layer  & $5^{th}$ Layer  & Parameters   \\
\hline
\hline
 Single MEG  & MLP-1  & FC (3000,'Sigmoid')   &  FC (2500,'Sigmoid') & FC (1200,'Sigmoid') & FC (3) & - & 11,428,303    \\
 Snapshot & MLP-2  & FC (3000,'Sigmoid')   &  FC (2500,'Sigmoid') & FC (1200,'Sigmoid') & FC (6) &-  &  11,431,906   \\
          & MLP-3  & FC (3000,'Sigmoid')   &  FC (2500,'Sigmoid') & FC (1200,'Sigmoid') & FC (9) &-  &  11,435,509   \\
\hline
MEG   & CNN-1  & Conv1D ($L=32$, $T=5$)   &   FC (3000,'Sigmoid')   &  FC (2500,'Sigmoid') & FC (1200,'Sigmoid') & FC (3) &   11,711,295\\
Time-Series             & CNN-2  & Conv1D ($L=32$, $T=5$)   & FC (3000,'Sigmoid')   &  FC (2500,'Sigmoid') & FC (1200,'Sigmoid') & FC (6)&  11,714,898\\
             & CNN-3  & Conv1D ($L=32$, $T=5$)   & FC (3000,'Sigmoid')   &  FC (2500,'Sigmoid') & FC (1200,'Sigmoid') & FC (9)& 11,718,501 \\

\hline
%MEG         & RNN-1     & Conv1D ($L=16$, $T=5$)      &  RNN (1024,'Tanh')     &  FC (1200,'Sigmoid')    & FC (3)   &  -      &  2,324,083 \\
%Time-Series & RNN-2     & Conv1D ($L=16$, $T=5$)      &  RNN (1024,'Tanh')     &  FC (1200,'Sigmoid')    & FC (6)   &   -     & 2,327,686\\
%            & RNN-3     & Conv1D ($L=16$, $T=5$)      &   RNN (1024,'Tanh')    &  FC (1200,'Sigmoid')    & FC (9)   &    -    & 2,331,289 \\ 
%\hline
\end{tabular}
\label{tab:Architectures_Summary}
\end{table*}
\subsection{Data Generation Workflow}

\begin{figure*}
  \centering
  \begin{tabular}[b]{c}
    \includegraphics[width=.25\linewidth]{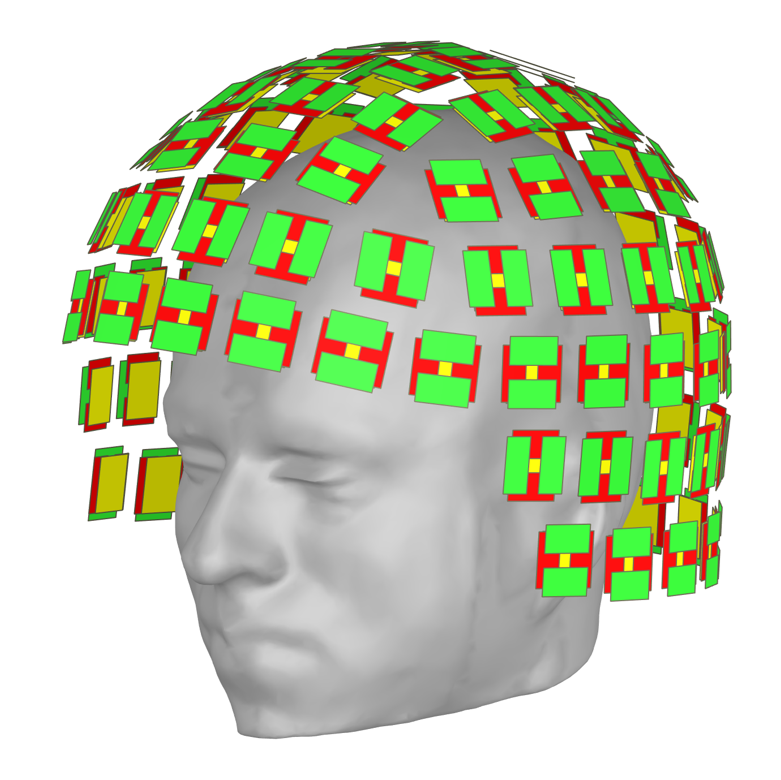} \\
    \small (a) MEG sensor geometry
  \end{tabular} \qquad
  \begin{tabular}[b]{c}
    \includegraphics[width=.30\linewidth]{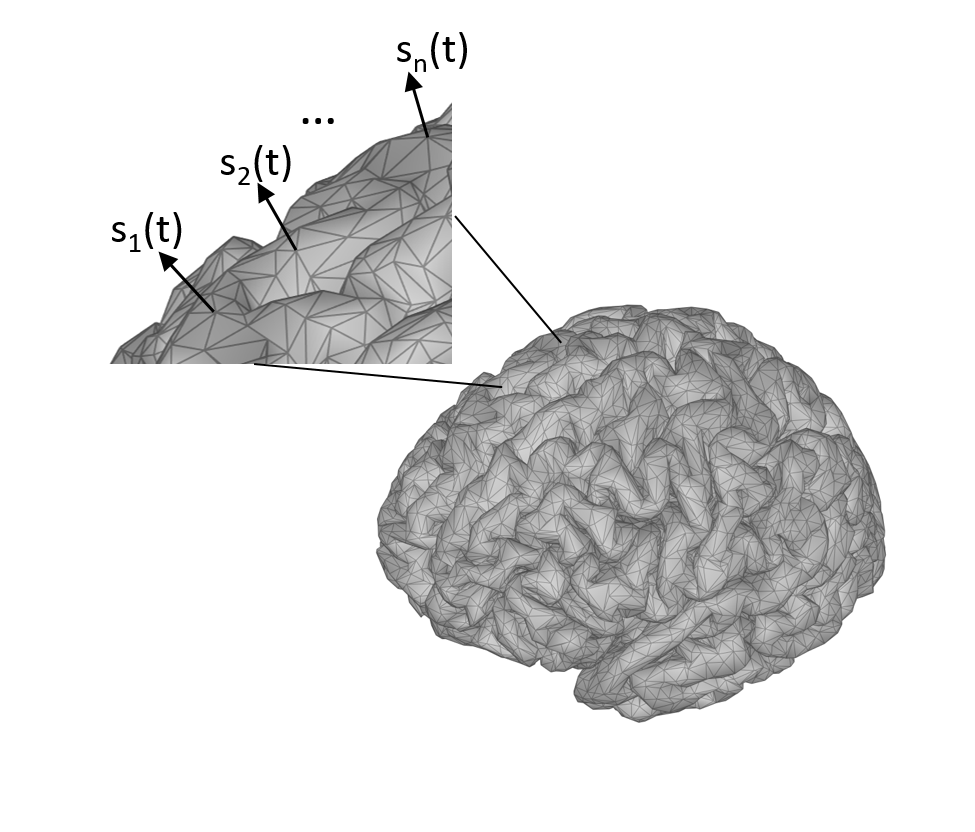} \\
    \small (b)  MEG source space
  \end{tabular} \qquad
  \begin{tabular}[b]{c}
    \includegraphics[width=.23\linewidth]{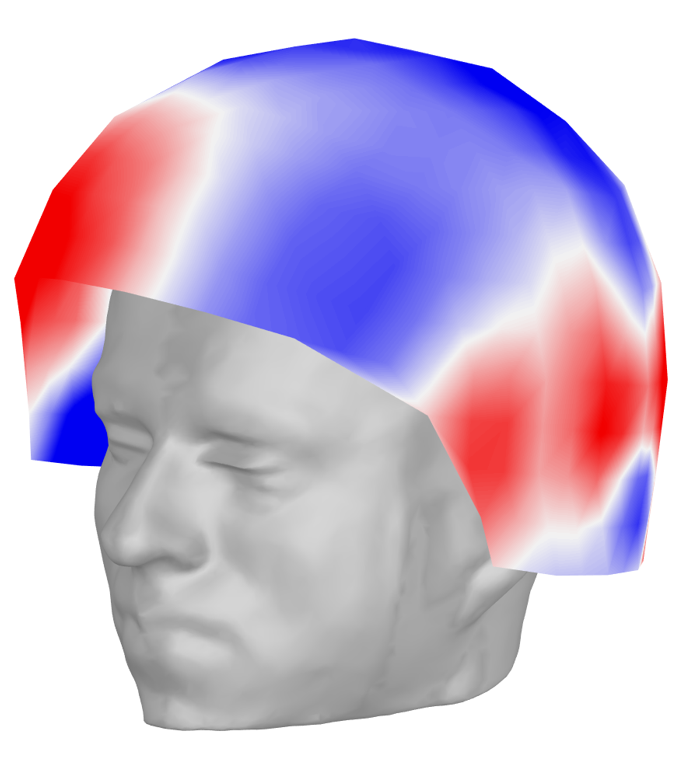} \\
    \small (c)  MEG simulated measurements
  \end{tabular}\\ 
\caption{Simulation of MEG data for deep learning localization. (a) Simulations used the anatomy of an adult human subject and a whole-head MEG sensor array from an Elekta Triux device. (b) Cortical sources with time course $s_i(t)$ were simulated at different cortical locations. (c) The activated cortical sources yielded MEG measurements on the cortex that, combined with additive Gaussian noise, comprised the input to the deep learning model. }
    \label{fig:data_generation}
\end{figure*}

To train the deep network models and evaluate their performance on source localization, we need to know the ground truth of the underlying neural sources generating MEG data. Since this information is unavailable in real MEG measurements of human participants, we performed simulations with an actual MEG sensor array and a realistic anatomy and source configurations. Specifically, the sensor array was based on the whole-head Elekta Triux MEG system (306-channel probe unit with 204 planar gradiometer sensors and 102 magnetometer sensors) (Figure \ref{fig:data_generation}a). The geometry of the MEG source space was modeled with the cortical manifold extracted from a T1-weighted MRI structural scan from a real adult human subject using Freesurfer \cite{fischl_2004}. This source configuration is consistent with the arrangement of pyramidal neurons, the principal source of MEG signals, in the cerebral cortex. Sources were restricted to approximately 15,000 grid points over the cortex (Figure \ref{fig:data_generation}b). The lead field matrix, which represents the forward mapping from the activated sources to the sensor array, was estimated using BrainStorm \cite{Brainstorm2011} based on an overlapping spheres head model \cite{Huang1999}. 

Simulated MEG sensor data was generated by first activating a few sources randomly selected on the cortical manifold with activation time courses $s_i(t)$. The time courses were modeled with 16 time points sampled as mixtures of sinusoidal signals. The corresponding sensor measurements were then obtained by multiplying each source with its respective topography vector $l(p_i)$ (Figure \ref{fig:data_generation}c). Finally, Gaussian white noise was generated and added to the MEG sensors to model instrumentation noise at specific SNR levels. The SNR was defined as the ratio of the Frobenius norm of the signal-magnetic-field spatiotemporal matrix to that of the noise matrix for each trial as in \cite{sekihara2001}.

\section{Performance Evaluation}

\subsection{Deep Network Training}

The DeepMEG models were implemented in TensorFlow \cite{tensorflow2015-whitepaper} and trained using the SGD algorithm (\ref{eq:SGD}) with a learning rate $\lambda=0.001$ and batch size of 32. The DeepMLP networks were trained with datasets of 1 million simulated snapshots, generated at a fixed SNR level, yet, as discussed in the following these trained models operate well in a wide range of SNR levels. The DeepCNN networks were trained using data generation on the fly\footnote{Data generation on the fly was utilized in order to mitigate the demanding memory requirements of offline data generation in the case of multiple-snapshots training set.} and a total of 9.6 million multiple-snapshot signals per network. The DeepCNN network models were trained with MEG sensor data at a fixed SNR-level and random inter-source correlations, thus, learning to localize sources with a wide range of inter-source correlation levels.

\subsection{Localization Experiments}
We assessed the performance of the deepMEG models using simulated data as described in the data generation workflow. To assess localization accuracy in different realistic scenarios, we conducted simulations with different SNR levels and inter-source correlation values.  We also varied the number of active sources to validate that localization is accurate even for multiple concurrently active sources. 

During inference, we compared the performance of the deep learning model against the popular scanning localization solution RAP-MUSIC \cite{Mosher1999}. All experiment were conducted with 1000 Monte-Carlo repetitions per each SNR and inter-source correlation value. In each scenario, we used the deepMEG and RAP-MUSIC with the corresponding number of sources, which are assumed known\footnote{Estimation of the number of sources can be conducted with the Akaike information criterion (AIC), Bayesian information criterion (BIC), or cross-validation and is beyond the scope of this work.} by both methods. 

\subsubsection{Experiment 1: Performance of the DeepMEG-MLP model with single-snapshot data}
\begin{figure}[ht!]
      \includegraphics[width=\columnwidth]{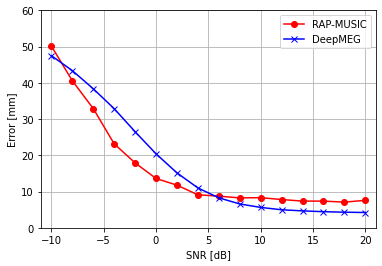} \\
    %\small (b)  Two Sources, Single Snapshot  
  \caption{Localization accuracy of the DeepMEG-MLP model at different SNR levels for the cases of two dipole sources.}
  \label{DeepMEG-MLPperformance}
\end{figure}

We assessed the localization accuracy of the DeepMEG-MLP model against the RAP-MUSIC method for the case of two simultaneously active dipole sources. The DeepMEG-MLP model was trained with 10 dB SNR data, but inference used different SNR levels ranging from -10 dB to 20 dB (Figure \ref{DeepMEG-MLPperformance}). The DeepMEG model outperformed the MAR-MUSIC method at high SNR values, but had worse localization results in low SNR values ($<5$ dB). As expected, both methods consistently improved their localization performance with increasing SNR values.

\subsubsection{Experiment 2: Performance of the DeepMEG-CNN model with multiple-snapshot data}
\begin{figure*}
  \centering
   
    \begin{tabular}[b]{c}
    \includegraphics[width=.45\linewidth]{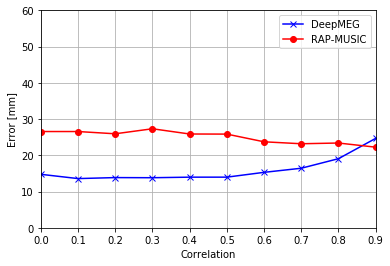} \\
    \small (a) Two Sources,  -15 dB SNR 
  \end{tabular} \qquad
    \begin{tabular}[b]{c}
    \includegraphics[width=.45\linewidth]{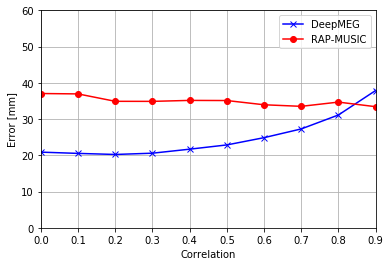} \\
    \small (b) Three Sources,  -15 dB SNR
  \end{tabular} \qquad

    \begin{tabular}[b]{c}
    \includegraphics[width=.45\linewidth]{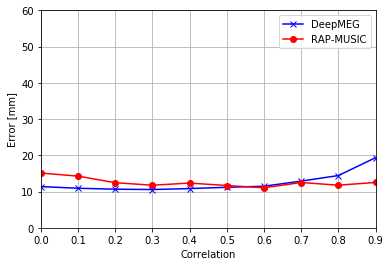} \\
    \small (c) Two Sources, -12.5 dB SNR 
  \end{tabular} \qquad
    \begin{tabular}[b]{c}
    \includegraphics[width=.45\linewidth]{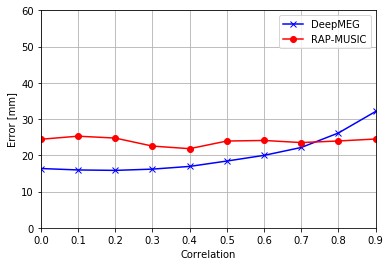} \\
    \small (d) Three Sources,  -12.5 dB SNR
   \end{tabular} \qquad
   
  \begin{tabular}[b]{c}
    \includegraphics[width=.45\linewidth]{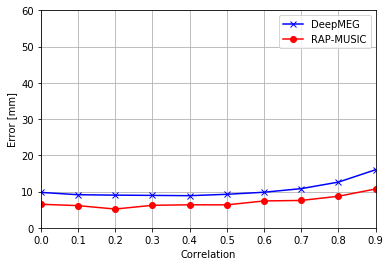} \\
    \small (e)  Two Sources, -10 dB SNR 
  \end{tabular} \qquad
     \begin{tabular}[b]{c}
    \includegraphics[width=.45\linewidth]{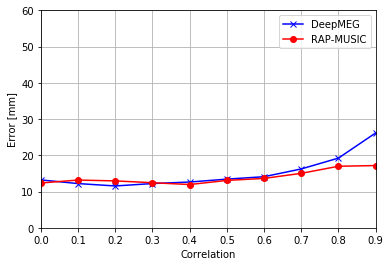} \\
    \small (f)  Three Sources,  -10 dB SNR
  \end{tabular} \qquad
  
   \caption{Localization accuracy of the DeepMEG-CNN model with $t_N=16$ snapshots at different inter-source correlation values for the cases of two and three sources with -15 dB, -12.5 dB and -10 dB SNR levels.}
  \label{DeepMEG-CNNperformance}
\end{figure*}

We extended the above experiment for the case of multiple snapshot data with $T=16$ time samples and two or three sources with different inter-source correlation values. The DeepMEG-CNN model was trained with -15 dB SNR data, and inference used -15 dB, -12.5 dB, and -10 dB SNR. In the low SNR case (-15 dB), the DeepMEG-CNN consistently outperformed the RAP-MUSIC method with the exception of high (0.9) correlation values where the RAP-MUSIC had a slightly better accuracy (Figure \ref{DeepMEG-CNNperformance}ab). As SNR increased to -12.5 dB, the Deep MEG model remained overall better or had comparable performance to RAP-MUSIC (Figure \ref{DeepMEG-CNNperformance}cd). This advantage was lost at -10 dB SNR, where RAP-MUSIC had an advantage (Figure \ref{DeepMEG-CNNperformance}ef).

\subsubsection{Experiment 3: Robustness of DeepMEG to forward model errors} 
\begin{figure*}
  \centering
  \begin{tabular}[b]{c}
    \includegraphics[width=.45\linewidth]{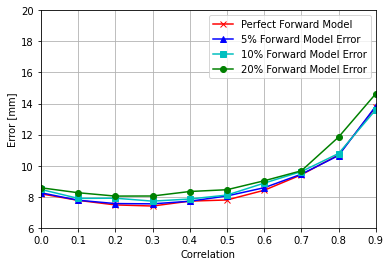} \\
    \small (a) Two sources 
  \end{tabular} \qquad
  \begin{tabular}[b]{c}
    \includegraphics[width=.45\linewidth]{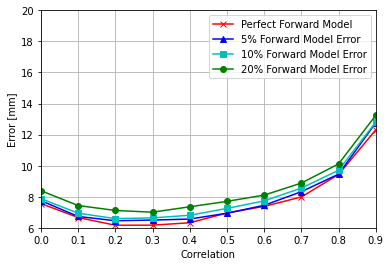} \\
    \small (b)  Three sources 
  \end{tabular}\\
   
  \caption{Robustness of DeepMEG-CNN localization accuracy to forward model errors ($T=16$ samples): (a) two sources at -5dB SNR. (b) three sources at 0dB SNR.}
  \label{fig:DL_robustness}
\end{figure*}

Here we assumed that the actual MEG forward model is different from the ideal forward model that was used for building the training set of the DeepMEG solution. The actual model was defined as follows: $\widetilde\mathbf A(\mathbf P) = \mathbf A(\mathbf P) +  \boldsymbol\Delta \mathbf A(\mathbf P)$, where  $\boldsymbol\Delta \mathbf{A}(\mathbf{P})$  denotes the  \textit{model error}, which can be \textit{arbitrary}. In this experiment we sampled the model error from a normal distribution, with zero mean and variance proportional to the Frobenius norm of the ideal forward model $\mathbf A(\mathbf P)$. Figure \ref{fig:DL_robustness} presents the results with model error variance equal to $5\%$, $10\%$, and $20\%$ of the Frobenius norm of $\mathbf A(\mathbf P)$, as compared to the performance without model error. The DeepMEG-CNN had robust localization accuracy in all cases.

\begin{comment}
\begin{figure*}[ht!]
  \centering
  \begin{tabular}[b]{c}
    \includegraphics[width=.45\linewidth]{figs/DeepMEG-1-SRC-1-SAMPLE.png} \\
    \small (a) One Source, Single Snapshot 
  \end{tabular} \qquad
  \begin{tabular}[b]{c}
    \includegraphics[width=.45\linewidth]{figs/RAP-MUSIC-DeepMEG-2-SRC-1-SAMPLE.png} \\
    \small (b)  Two Sources, Single Snapshot  
  \end{tabular}\\
  \caption{Localization accuracy of the DeepMEG-MLP model at different SNR levels for the cases of one and two dipole sources.}
  \label{DeepMEG-MLPperformance}
\end{figure*}
\end{comment}

\subsection{Real-Time Source Localization}
An important advantage of the DL approaches is that they have significantly reduced computational time, paving the way to real-time MEG source localization solutions. We conducted a computation time comparison\footnote{Compute hardware: CPU 6-Core Intel i5-9400F@3.9Ghz, CPU RAM 64GB, GPU Nvidia GeForce RTX 2080.}, detailed in table \ref{tab:Comp_Time}, for each of the proposed DL architectures and the RAP-MUSIC algorithm. The comparison reveals 4 orders of magnitude faster computation of DeepMEG models as compared to the RAP-MUSIC, which requires $O(Q)$ $M\times M$ matrix inversions and $O(Q \times N_{dipoles})$ matrix-vector multiplications, with each multiplication having complexity of $O(M^2)$. Here, the total number of dipoles $N_{dipoles}=15,002$ and the number of sensors $ M=306$.   

\begin{table}
\caption{Computation Time Comparison }
\centering
\begin{tabular}{|c|c|c|l|}
\hline
Sources & Time Samples &  Algorithm & Time [ms] \\
\hline
\hline
1       & 1            &  RAP-MUSIC   & 135.47      \\
%1       & 1            &  RAP-BeamFormer   & 197,521.62       \\       
1       & 1            &  DeepMEG MLP-1   &  \textbf{0.19}      \\   
\hline
2       & 1            &  RAP-MUSIC   &  452.17   \\
%2       & 1            &  RAP-BeamFormer   &  552,063.56        \\       
2       & 1            &  DeepMEG MLP-2   & \textbf{0.19}       \\
\hline
3       & 1            &  RAP-MUSIC   &  736.76 \\
%3       & 1            &  RAP-BeamFormer   & 916,120.15       \\       
3       & 1            &  DeepMEG MLP-3   & \textbf{0.19}       \\
\hline
1       & 16            &  RAP-MUSIC   &    136.59  \\
%1       & 16            &  RAP-BeamFormer   & 203,046.78       \\       
1       & 16            &  DeepMEG CNN-1   & \textbf{0.25}      \\ 
%1       & 16            &  DeepMEG RNN-1   & \textbf{1.08}      \\   
\hline
2       & 16            &  RAP-MUSIC   &   478.23  \\
%2       & 16            &  RAP-BeamFormer   & 552,981.40       \\       
2       & 16            &  DeepMEG CNN-2   & \textbf{0.27}       \\ 
%2       & 16            &  DeepMEG RNN-2   & \textbf{1.09}      \\
\hline
3       & 16            &  RAP-MUSIC   &   741.51 \\
%3       & 16            &  RAP-BeamFormer   & 912,789.08       \\       
3       & 16            &  DeepMEG CNN-3   & \textbf{0.27}     \\ 
%3       & 16            &  DeepMEG RNN-3   & \textbf{1.16}     \\
\hline
\end{tabular}
\label{tab:Comp_Time}
\end{table}

\section{Conclusions}

Fast and accurate solutions to MEG source localization are crucial for real-time brain imaging, and hold the potential to enable novel applications in neurorehabilitation and BMI. Current scanning solutions have low resolution and limit the number of dipoles and temporal rate of source localization due to high computational demands. In this article, we reviewed existing MEG source localization solutions and fundamental DL tools. Motivated by the recent success of DL in a growing number of inverse imaging problems, we proposed two DL architectures for the solution of the MEG inverse problem, the DeepMEG-MLP for single time point localization, and the DeepMEG-CNN for multiple time point localization. 

We compared the performance of DeepMEG against the popular RAP-MUSIC localization algorithm and showed improvements in localization accuracy in a range of scenarios with variable SNR levels, inter-source correlation values, and number of sources. Importantly, the DeepMEG inference was estimable in less than a millisecond and thus was orders of magnitude faster than RAP-MUSIC. Fast computation was possible due to the high optimization of modern DL tools, and even allows the rapid estimation of dipoles at near the 1 kHz sampling rate speed of existing MEG devices. This could facilitate the search for optimal indices of brain activity in neurofeedback and BCI tasks. 

A key property of the DeepMEG model was its robustness to forward model errors. The localization performance of the model remained stable even when distortions were up to $20\%$ of the Frobenious norm of the lead field matrix. This is critical for real-time applications where the forward matrix is not precisely known, or movement of the subject introduces time-varying inaccuracies.

While the DeepMEG-MLP and DeepMEG-CNN architectures yielded promising localization results, future work is needed to explore different architectures, regularizations, loss functions, and other DL parameters that may further improve MEG source localization. This is particularly important in MEG and EEG source localization where DL tools have not yet been developed, with the exception of a hybrid dSPM-LSTM solution \cite{dinh2019contextual}.

\section{Authors}
\textbf{Dimitrios Pantazis} is a Principal Research Scientist and the Director of the MEG laboratory at the McGovern Institute for Brain Research, Massachusetts Institute of Technology. Before joining MIT, he was Research Assistant Professor at the University of Southern California from 2008-2010. He received his PhD in Electrical Engineering at the University of Southern California in 2006. He has 20 years of experience in developing methods for the analysis of MEG data and has published prominent articles in Nature Neuroscience, Nature Communications, Proceedings of the National Academy of Sciences, Scientific Reports, Cerebral Cortex, NeuroImage, and others. He is a key developer of Brainstorm \cite{Brainstorm2011}, an open-source environment dedicated to the analysis of brain recordings (MEG, EEG, NIRS, ECoG, depth electrodes, animal electrophysiology) with 13,000+ registered users and 400+ related publications. His research focuses on novel methodology for resolving neural representations from MEG data, development of multimodal imaging techniques, and characterization of pathological function in neurological disorders.\\

\textbf{Amir Adler} is a Senior Lecturer at the Electrical Engineering Department, ORT Braude College of Engineering, Israel, and a Research Affiliate at the McGovern Institute for Brain Research, MIT. Previously, he was a post-doctoral associate at the Center for Brains, Minds, and Machines (CBMM), MIT, where his research was focused on deep learning and inverse problems. He served as the Chief Scientist of the Ministry of Communications, Israel, and was Chief Technology Officer of the Wi-Fi division at NextWave Wireless Inc. (acquired by AT\&T). He is the recipient of the 2011 Google Europe Doctoral Fellowship in Multimedia, a co-recipient of the 2012 HP Labs Open Innovation Research Award, and a recipient of the 2019 Facebook Research Content Policy Research award\footnote{\url{https://research.fb.com/blog/2019/05/announcing-the-winners-of-the-content-policy-research-on-social-media-platforms-research-awards/}}, for his research on deep generative models. His current research is focused in deep learning, inverse problems and source localization. He holds a B.Sc. (Cum Laude) and M.Eng. (Cum Laude) in Electrical Engineering, and a Ph.D. in Computer Science, all from the Technion - Israel Institute of Technology. \\

%\section{Introduction}
%\label{sec:seismic-background}
%\input{background.tex}
%

%
%
%\section{Subsurface Model Building}
%\label{sec:tomo}
%\input{tomo.tex}
%

%
%\section{The Deep Learning approach}
%\label{sec:dl}
%\input{dl.tex}

% if have a single appendix:
%\appendix[Proof of the Zonklar Equations]
% or
%\appendix  % for no appendix heading
% do not use \section anymore after \appendix, only \section*
% is possibly needed

% use appendices with more than one appendix
% then use \section to start each appendix
% you must declare a \section before using any
% \subsection or using \label (\appendices by itself
% starts a section numbered zero.)
%

%\appendices
%\section{Proof of the First Zonklar Equation}
%Appendix one text goes here.

% you can choose not to have a title for an appendix
% if you want by leaving the argument blank
%\section{}
%Appendix two text goes here.

% use section* for acknowledgment

% Can use something like this to put references on a page
% by themselves when using endfloat and the captionsoff option.
\ifCLASSOPTIONcaptionsoff
  \newpage
\fi

% trigger a \newpage just before the given reference
% number - used to balance the columns on the last page
% adjust value as needed - may need to be readjusted if
% the document is modified later
%\IEEEtriggeratref{8}
% The "triggered" command can be changed if desired:
%\IEEEtriggercmd{\enlargethispage{-5in}}

% references section

% can use a bibliography generated by BibTeX as a .bbl file
% BibTeX documentation can be easily obtained at:
% http://mirror.ctan.org/biblio/bibtex/contrib/doc/
% The IEEEtran BibTeX style support page is at:
% http://www.michaelshell.org/tex/ieeetran/bibtex/
%\bibliographystyle{IEEEtran}
% argument is your BibTeX string definitions and bibliography database(s)
%\bibliography{IEEEabrv,../bib/paper}
%
% <OR> manually copy in the resultant .bbl file
% set second argument of \begin to the number of references
% (used to reserve space for the reference number labels box)
\bibliographystyle{IEEEtran}
\bibliography{references}

% Generated by IEEEtran.bst, version: 1.14 (2015/08/26)
\begin{thebibliography}{10}
\providecommand{\url}[1]{#1}
\csname url@samestyle\endcsname
\providecommand{\newblock}{\relax}
\providecommand{\bibinfo}[2]{#2}
\providecommand{\BIBentrySTDinterwordspacing}{\spaceskip=0pt\relax}
\providecommand{\BIBentryALTinterwordstretchfactor}{4}
\providecommand{\BIBentryALTinterwordspacing}{\spaceskip=\fontdimen2\font plus
\BIBentryALTinterwordstretchfactor\fontdimen3\font minus
  \fontdimen4\font\relax}
\providecommand{\BIBforeignlanguage}[2]{{%
\expandafter\ifx\csname l@#1\endcsname\relax
\typeout{** WARNING: IEEEtran.bst: No hyphenation pattern has been}%
\typeout{** loaded for the language `#1'. Using the pattern for}%
\typeout{** the default language instead.}%
\else
\language=\csname l@#1\endcsname
\fi
#2}}
\providecommand{\BIBdecl}{\relax}
\BIBdecl

\bibitem{VALEROCABRE2017381}
A.~Valero-Cabré, J.~L. Amengual, C.~Stengel, A.~Pascual-Leone, and O.~A.
  Coubard, ``Transcranial magnetic stimulation in basic and clinical
  neuroscience: A comprehensive review of fundamental principles and novel
  insights,'' \emph{Neuroscience and Biobehavioral Reviews}, vol.~83, pp. 381
  -- 404, 2017.

\bibitem{KOHL2019355}
S.~Kohl, R.~Hannah, L.~Rocchi, C.~L. Nord, J.~Rothwell, and V.~Voon, ``Cortical
  paired associative stimulation influences response inhibition:
  Cortico-cortical and cortico-subcortical networks,'' \emph{Biological
  Psychiatry}, vol.~85, no.~4, pp. 355 -- 363, 2019.

\bibitem{FOLLONI20191109}
D.~Folloni, L.~Verhagen, R.~B. Mars, E.~Fouragnan, C.~Constans, J.-F. Aubry,
  M.~F. Rushworth, and J.~Sallet, ``Manipulation of subcortical and deep
  cortical activity in the primate brain using transcranial focused ultrasound
  stimulation,'' \emph{Neuron}, vol. 101, no.~6, pp. 1109 -- 1116.e5, 2019.

\bibitem{Darrow2019}
D.~Darrow, ``Focused ultrasound for neuromodulation,''
  \emph{Neurotherapeutics}, vol.~16, no.~6, pp. 88 -- 99, 2019.

\bibitem{MIN2017585}
B.-K. Min, R.~Chavarriaga, and J.~del R.~Millán, ``Harnessing prefrontal
  cognitive signals for brain–machine interfaces,'' \emph{Trends in
  Biotechnology}, vol.~35, no.~7, pp. 585 -- 597, 2017.

\bibitem{PICHIORRI2020101}
\BIBentryALTinterwordspacing
F.~Pichiorri and D.~Mattia, ``Brain-computer interfaces in neurologic
  rehabilitation practice,'' in \emph{Brain-Computer Interfaces}, ser. Handbook
  of Clinical Neurology.\hskip 1em plus 0.5em minus 0.4em\relax Elsevier, 2020,
  vol. 168, pp. 101 -- 116. [Online]. Available:
  \url{http://www.sciencedirect.com/science/article/pii/B9780444639349000093}
\BIBentrySTDinterwordspacing

\bibitem{ilmoniemi2019brain}
\BIBentryALTinterwordspacing
R.~Ilmoniemi and J.~Sarvas, \emph{Brain Signals: Physics and Mathematics of MEG
  and EEG}.\hskip 1em plus 0.5em minus 0.4em\relax MIT Press, 2019. [Online].
  Available: \url{https://books.google.co.il/books?id=y5iWDwAAQBAJ}
\BIBentrySTDinterwordspacing

\bibitem{niedermeyer_niedermeyers_2012}
E.~Niedermeyer, D.~L. Schomer, and F.~H. Lopes~da Silva, \emph{Niedermeyer's
  electroencephalography: basic principles, clinical applications, and related
  fields}, 2012, {OCLC}: 1062345900.

\bibitem{hamalainen_magnetoencephalographytheory_1993}
\BIBentryALTinterwordspacing
M.~Hämäläinen, R.~Hari, R.~J. Ilmoniemi, J.~Knuutila, and O.~V. Lounasmaa,
  ``\BIBforeignlanguage{en}{Magnetoencephalography—theory, instrumentation,
  and applications to noninvasive studies of the working human brain},''
  \emph{\BIBforeignlanguage{en}{Reviews of Modern Physics}}, vol.~65, no.~2,
  pp. 413--497, Apr. 1993. [Online]. Available:
  \url{http://link.aps.org/doi/10.1103/RevModPhys.65.413}
\BIBentrySTDinterwordspacing

\bibitem{baillet_magnetoencephalography_2017}
S.~Baillet, ``Magnetoencephalography for brain electrophysiology and imaging,''
  \emph{Nature Neuroscience}, vol.~20, no.~3, pp. 327--339, 2017.

\bibitem{DARVAS2004S289}
F.~Darvas, D.~Pantazis, E.~Kucukaltun-Yildirim, and R.~Leahy, ``Mapping human
  brain function with {MEG} and {EEG}: methods and validation,''
  \emph{NeuroImage}, vol.~23, pp. S289 -- S299, 2004.

\bibitem{kleiner_superconducting_2004}
\BIBentryALTinterwordspacing
R.~Kleiner, D.~Koelle, F.~Ludwig, and J.~Clarke, ``Superconducting quantum
  interference devices: State of the art and applications,'' \emph{Proceedings
  of the {IEEE}}, vol.~92, no.~10, pp. 1534--1548, 2004. [Online]. Available:
  \url{http://ieeexplore.ieee.org/document/1335547/}
\BIBentrySTDinterwordspacing

\bibitem{Goodfellow-et-al-2016}
I.~Goodfellow, Y.~Bengio, and A.~Courville, \emph{Deep Learning}.\hskip 1em
  plus 0.5em minus 0.4em\relax MIT Press, 2016,
  \url{http://www.deeplearningbook.org}.

\bibitem{gupta_cnn-based_2018}
\BIBentryALTinterwordspacing
H.~Gupta, K.~H. Jin, H.~Q. Nguyen, M.~T. {McCann}, and M.~Unser, ``{CNN}-based
  projected gradient descent for consistent {CT} image reconstruction,''
  \emph{{IEEE} Transactions on Medical Imaging}, vol.~37, no.~6, pp.
  1440--1453, 2018. [Online]. Available:
  \url{https://ieeexplore.ieee.org/document/8353870/}
\BIBentrySTDinterwordspacing

\bibitem{jin_deep_2017}
\BIBentryALTinterwordspacing
K.~H. Jin, M.~T. {McCann}, E.~Froustey, and M.~Unser, ``Deep convolutional
  neural network for inverse problems in imaging,'' \emph{{IEEE} Transactions
  on Image Processing}, vol.~26, no.~9, pp. 4509--4522, 2017. [Online].
  Available: \url{http://ieeexplore.ieee.org/document/7949028/}
\BIBentrySTDinterwordspacing

\bibitem{flohr_deep_2017}
\BIBentryALTinterwordspacing
L.~Gjesteby, Q.~Yang, Y.~Xi, Y.~Zhou, J.~Zhang, and G.~Wang, ``Deep learning
  methods to guide {CT} image reconstruction and reduce metal artifacts,''
  T.~G. Flohr, J.~Y. Lo, and T.~Gilat~Schmidt, Eds., 2017, p. 101322W.
  [Online]. Available:
  \url{http://proceedings.spiedigitallibrary.org/proceeding.aspx?doi=10.1117/12.2254091}
\BIBentrySTDinterwordspacing

\bibitem{8962949}
D.~{Liang}, J.~{Cheng}, Z.~{Ke}, and L.~{Ying}, ``Deep magnetic resonance image
  reconstruction: Inverse problems meet neural networks,'' \emph{IEEE Signal
  Processing Magazine}, vol.~37, no.~1, pp. 141--151, Jan 2020.

\bibitem{wang_accelerating_2016}
\BIBentryALTinterwordspacing
S.~{Wang}, Z.~{Su}, L.~{Ying}, X.~{Peng}, S.~{Zhu}, F.~{Liang}, D.~{Feng}, and
  D.~{Liang}, ``Accelerating magnetic resonance imaging via deep learning,'' in
  \emph{2016 IEEE 13th International Symposium on Biomedical Imaging (ISBI)},
  2016. [Online]. Available: \url{http://ieeexplore.ieee.org/document/7493320/}
\BIBentrySTDinterwordspacing

\bibitem{schlemper_deep_2018}
\BIBentryALTinterwordspacing
J.~Schlemper, J.~Caballero, J.~V. Hajnal, A.~N. Price, and D.~Rueckert, ``A
  deep cascade of convolutional neural networks for dynamic {MR} image
  reconstruction,'' \emph{{IEEE} Transactions on Medical Imaging}, vol.~37,
  no.~2, pp. 491--503, 2018. [Online]. Available:
  \url{https://ieeexplore.ieee.org/document/8067520/}
\BIBentrySTDinterwordspacing

\bibitem{gong_pet_2018}
\BIBentryALTinterwordspacing
K.~Gong, C.~Catana, J.~Qi, and Q.~Li, ``{PET} image reconstruction using deep
  image prior,'' \emph{{IEEE} Transactions on Medical Imaging}, pp. 1--1, 2018.
  [Online]. Available: \url{https://ieeexplore.ieee.org/document/8581448/}
\BIBentrySTDinterwordspacing

\bibitem{kim_penalized_2018}
\BIBentryALTinterwordspacing
K.~Kim, D.~Wu, K.~Gong, J.~Dutta, J.~H. Kim, Y.~D. Son, H.~K. Kim,
  G.~El~Fakhri, and Q.~Li, ``Penalized {PET} reconstruction using deep learning
  prior and local linear fitting,'' \emph{{IEEE} Transactions on Medical
  Imaging}, vol.~37, no.~6, pp. 1478--1487, 2018. [Online]. Available:
  \url{https://ieeexplore.ieee.org/document/8354909/}
\BIBentrySTDinterwordspacing

\bibitem{dong_image_2016}
\BIBentryALTinterwordspacing
C.~Dong, C.~C. Loy, K.~He, and X.~Tang, ``Image super-resolution using deep
  convolutional networks,'' \emph{{IEEE} Transactions on Pattern Analysis and
  Machine Intelligence}, vol.~38, no.~2, pp. 295--307, 2016. [Online].
  Available: \url{http://ieeexplore.ieee.org/document/7115171/}
\BIBentrySTDinterwordspacing

\bibitem{kim_accurate_2016}
\BIBentryALTinterwordspacing
J.~Kim, J.~K. Lee, and K.~M. Lee, ``Accurate image super-resolution using very
  deep convolutional networks,'' in \emph{2016 {IEEE} Conference on Computer
  Vision and Pattern Recognition ({CVPR})}.\hskip 1em plus 0.5em minus
  0.4em\relax {IEEE}, 2016, pp. 1646--1654. [Online]. Available:
  \url{http://ieeexplore.ieee.org/document/7780551/}
\BIBentrySTDinterwordspacing

\bibitem{lim_enhanced_2017}
\BIBentryALTinterwordspacing
B.~Lim, S.~Son, H.~Kim, S.~Nah, and K.~M. Lee, ``Enhanced deep residual
  networks for single image super-resolution,'' in \emph{2017 {IEEE} Conference
  on Computer Vision and Pattern Recognition Workshops ({CVPRW})}.\hskip 1em
  plus 0.5em minus 0.4em\relax {IEEE}, 2017, pp. 1132--1140. [Online].
  Available: \url{http://ieeexplore.ieee.org/document/8014885/}
\BIBentrySTDinterwordspacing

\bibitem{hauptmann_model-based_2018}
A.~Hauptmann, F.~Lucka, M.~Betcke, N.~Huynh, J.~Adler, B.~Cox, P.~Beard,
  S.~Ourselin, and S.~Arridge, ``Model-based learning for accelerated,
  limited-view 3-d photoacoustic tomography,'' \emph{{IEEE} Transactions on
  Medical Imaging}, vol.~37, no.~6, pp. 1382--1393, 2018.

\bibitem{yonel_deep_2018}
\BIBentryALTinterwordspacing
B.~Yonel, E.~Mason, and B.~Yazici, ``Deep learning for passive synthetic
  aperture radar,'' \emph{{IEEE} Journal of Selected Topics in Signal
  Processing}, vol.~12, no.~1, pp. 90--103, 2018. [Online]. Available:
  \url{https://ieeexplore.ieee.org/document/8214209/}
\BIBentrySTDinterwordspacing

\bibitem{budillon_sar_2019}
\BIBentryALTinterwordspacing
A.~Budillon, A.~C. Johnsy, G.~Schirinzi, and S.~Vitale, ``Sar tomography based
  on deep learning,'' in \emph{{IGARSS} 2019 - 2019 {IEEE} International
  Geoscience and Remote Sensing Symposium}.\hskip 1em plus 0.5em minus
  0.4em\relax {IEEE}, 2019, pp. 3625--3628. [Online]. Available:
  \url{https://ieeexplore.ieee.org/document/8900616/}
\BIBentrySTDinterwordspacing

\bibitem{Araya2018}
\BIBentryALTinterwordspacing
M.~Araya-Polo, J.~Jennings, A.~Adler, and T.~Dahlke, ``Deep-learning
  tomography,'' \emph{The Leading Edge}, vol.~37, no.~1, pp. 58--66, 2018.
  [Online]. Available: \url{https://doi.org/10.1190/tle37010058.1}
\BIBentrySTDinterwordspacing

\bibitem{mosher_multiple_1992}
\BIBentryALTinterwordspacing
J.~Mosher, P.~Lewis, and R.~Leahy, ``\BIBforeignlanguage{en}{Multiple dipole
  modeling and localization from spatio-temporal {MEG} data},''
  \emph{\BIBforeignlanguage{en}{IEEE Transactions on Biomedical Engineering}},
  vol.~39, no.~6, pp. 541--557, Jun. 1992. [Online]. Available:
  \url{http://ieeexplore.ieee.org/document/141192/}
\BIBentrySTDinterwordspacing

\bibitem{huang_multi-start_1998}
\BIBentryALTinterwordspacing
M.~Huang, C.~Aine, S.~Supek, E.~Best, D.~Ranken, and E.~Flynn,
  ``\BIBforeignlanguage{en}{Multi-start downhill simplex method for
  spatio-temporal source localization in magnetoencephalography},''
  \emph{\BIBforeignlanguage{en}{Electroencephalography and Clinical
  Neurophysiology/Evoked Potentials Section}}, vol. 108, no.~1, pp. 32--44,
  Jan. 1998. [Online]. Available:
  \url{https://linkinghub.elsevier.com/retrieve/pii/S0168559797000919}
\BIBentrySTDinterwordspacing

\bibitem{uutela_global_1998}
\BIBentryALTinterwordspacing
K.~Uutela, M.~Hamalainen, and R.~Salmelin, ``\BIBforeignlanguage{en}{Global
  optimization in the localization of neuromagnetic sources},''
  \emph{\BIBforeignlanguage{en}{IEEE Transactions on Biomedical Engineering}},
  vol.~45, no.~6, pp. 716--723, Jun. 1998. [Online]. Available:
  \url{http://ieeexplore.ieee.org/document/678606/}
\BIBentrySTDinterwordspacing

\bibitem{khosla_spatio-temporal_1997}
\BIBentryALTinterwordspacing
D.~Khosla, M.~Singh, and M.~Don, ``\BIBforeignlanguage{en}{Spatio-temporal
  {EEG} source localization using simulated annealing},''
  \emph{\BIBforeignlanguage{en}{IEEE Transactions on Biomedical Engineering}},
  vol.~44, no.~11, pp. 1075--1091, Nov. 1997. [Online]. Available:
  \url{http://ieeexplore.ieee.org/document/641335/}
\BIBentrySTDinterwordspacing

\bibitem{jiang_comparative_2003}
\BIBentryALTinterwordspacing
T.~Jiang, A.~Luo, X.~Li, and F.~Kruggel, ``\BIBforeignlanguage{en}{A
  {Comparative} {Study} {Of} {Global} {Optimization} {Approaches} {To} {Meg}
  {Source} {Localization}},'' \emph{\BIBforeignlanguage{en}{International
  Journal of Computer Mathematics}}, vol.~80, no.~3, pp. 305--324, Mar. 2003.
  [Online]. Available:
  \url{http://www.tandfonline.com/doi/abs/10.1080/0020716022000009255}
\BIBentrySTDinterwordspacing

\bibitem{darvas_mapping_2004}
\BIBentryALTinterwordspacing
F.~Darvas, D.~Pantazis, E.~Kucukaltun-Yildirim, and R.~Leahy,
  ``\BIBforeignlanguage{en}{Mapping human brain function with {MEG} and {EEG}:
  methods and validation},'' \emph{\BIBforeignlanguage{en}{NeuroImage}},
  vol.~23, pp. S289--S299, Jan. 2004. [Online]. Available:
  \url{http://linkinghub.elsevier.com/retrieve/pii/S1053811904003799}
\BIBentrySTDinterwordspacing

\bibitem{LCMV}
{B. D. van Veen, W. van Drongelen, M. Yuchtman, and A. Suzuki}, ``Localization
  of brain electrical activity via linearly constrained minimum variance
  spatial filtering,'' \emph{IEEE Trans. Biomed. Eng.}, vol.~44, no.~9, pp.
  867--880, Sep 1997.

\bibitem{VerbaRobinson}
J.~Vrba and S.~E. Robinson, ``Signal processing in magnetoencephalography,''
  \emph{Methods}, vol.~25, pp. 249--271, 2001.

\bibitem{CohBF}
{S. S. {Dalal} and K. {Sekihara} and S. S. {Nagarajan}}, ``Modified beamformers
  for coherent source region suppression,'' \emph{IEEE Transactions on
  Biomedical Engineering}, vol.~53, no.~7, pp. 1357--1363, July 2006.

\bibitem{CorrBF}
{M. J. {Brookes}, C. M. {Stevenson}, G. R. {Barnes}, A. {Hillebrand}, M. I.
  {Simpson}, S. T. {Francis}, and P. G. {Morris}}, ``Beamformer reconstruction
  of correlated sources using a modified source,'' \emph{NeuroImage}, vol.~34,
  no.~4, pp. 1454--1465, 2007.

\bibitem{NullBF}
{H. B. Hu, D. Pantazis, S. L. Bressler and R. M. Leahy}, ``Identifying true
  cortical interactions in {MEG} using the nulling beamformer,''
  \emph{NeuroImage}, vol.~49, no.~4, pp. 3161--3174, 2010.

\bibitem{DCBF}
{M. Diwakar and M.X. Huang and R. Srinivasan and D. L. Harrington and A. Robb
  and A. Angeles and L. Muzzatti and R. Pakdaman and T. Song and R. J.
  Theilmann and R. R. Lee, }, ``Dual-core beamformer for obtaining highly
  correlated neuronal networks in meg,'' \emph{NeuroImage}, vol.~54, no.~1, pp.
  253--263, 2011.

\bibitem{EDCBF}
M.~Diwakar, O.~Tal, T.~Liu, D.~Harrington, R.~Srinivasan, L.~Muzzatti, T.~Song,
  R.~Theilmann, R.~Lee, and M.~Huang, ``Accurate reconstruction of temporal
  correlation for neuronal sources using the enhanced dual-core meg
  beamformer,'' \emph{NeuroImage}, vol.~56, pp. 1918--1928, 2011.

\bibitem{MultiLCMV}
{A. Moiseev, J. M. Gaspar, J. A. Schneider and A. T. Herdman}, ``Application of
  multi-source minimum variance beamformers for reconstruction of correlated
  neural activity,'' \emph{NeuroImage}, vol.~58, no.~2, pp. 481--496, Sep 2011.

\bibitem{MultiBeamformers}
A.~{Moiseev} and A.~T. {Herdman}, ``Multi-core beamformers: Derivation,
  limitations and improvements,'' \emph{NeuroImage}, vol.~71, pp. 135--146,
  2013.

\bibitem{POP-MUSIC}
{Hesheng Liu} and P.~H. {Schimpf}, ``Efficient localization of synchronous eeg
  source activities using a modified rap-music algorithm,'' \emph{IEEE
  Transactions on Biomedical Engineering}, vol.~53, no.~4, pp. 652--661, April
  2006.

\bibitem{WedgeMUSIC}
{A. Ewald, F. S. Avarvand and G. Nolte}, ``Wedge {MUSIC}: A novel approach to
  examine experimental differences of brain source connectivity patterns from
  {EEG/MEG} data,'' \emph{NeuroImage}, vol. 101, pp. 610--624, 2014.

\bibitem{hui_identifying_2010}
\BIBentryALTinterwordspacing
H.~B. Hui, D.~Pantazis, S.~L. Bressler, and R.~M. Leahy,
  ``\BIBforeignlanguage{en}{Identifying true cortical interactions in {MEG}
  using the nulling beamformer},'' \emph{\BIBforeignlanguage{en}{NeuroImage}},
  vol.~49, no.~4, pp. 3161--3174, Feb. 2010. [Online]. Available:
  \url{http://linkinghub.elsevier.com/retrieve/pii/S1053811909011501}
\BIBentrySTDinterwordspacing

\bibitem{Mosher1999}
J.~C. {Mosher} and R.~M. {Leahy}, ``Source localization using recursively
  applied and projected {(RAP) MUSIC},'' \emph{IEEE Transactions on Signal
  Processing}, vol.~47, no.~2, pp. 332--340, Feb 1999.

\bibitem{makela_truncated_2018}
\BIBentryALTinterwordspacing
N.~Mäkelä, M.~Stenroos, J.~Sarvas, and R.~J. Ilmoniemi,
  ``\BIBforeignlanguage{en}{Truncated {RAP}-{MUSIC} ({TRAP}-{MUSIC}) for {MEG}
  and {EEG} source localization},'' \emph{\BIBforeignlanguage{en}{NeuroImage}},
  vol. 167, pp. 73--83, Feb. 2018. [Online]. Available:
  \url{https://linkinghub.elsevier.com/retrieve/pii/S1053811917309205}
\BIBentrySTDinterwordspacing

\bibitem{makela_locating_2017}
\BIBentryALTinterwordspacing
------, ``\BIBforeignlanguage{en}{Locating highly correlated sources from {MEG}
  with (recursive) ({R}){DS}-{MUSIC}},'' Neuroscience, preprint, Dec. 2017.
  [Online]. Available: \url{http://biorxiv.org/lookup/doi/10.1101/230672}
\BIBentrySTDinterwordspacing

\bibitem{8253590}
A.~Lucas, M.~Iliadis, R.~Molina, and A.~K. Katsaggelos, ``Using deep neural
  networks for inverse problems in imaging: Beyond analytical methods,''
  \emph{IEEE Signal Processing Magazine}, vol.~35, no.~1, pp. 20--36, Jan 2018.

\bibitem{Deep_MRI_8962949}
D.~{Liang}, J.~{Cheng}, Z.~{Ke}, and L.~{Ying}, ``Deep magnetic resonance image
  reconstruction: Inverse problems meet neural networks,'' \emph{IEEE Signal
  Processing Magazine}, vol.~37, no.~1, pp. 141--151, 2020.

\bibitem{6472238}
Y.~{Bengio}, A.~{Courville}, and P.~{Vincent}, ``Representation learning: A
  review and new perspectives,'' \emph{IEEE Transactions on Pattern Analysis
  and Machine Intelligence}, vol.~35, no.~8, pp. 1798--1828, 2013.

\bibitem{CAE}
J.~Masci, U.~Meier, D.~Cire\c{s}an, and J.~Schmidhuber, ``Stacked convolutional
  auto-encoders for hierarchical feature extraction,'' in \emph{Proceedings of
  the 21th International Conference on Artificial Neural Networks
  (ICANN’11)}.\hskip 1em plus 0.5em minus 0.4em\relax Springer-Verlag, 2011,
  p. 52–59.

\bibitem{RFB15a}
O.~Ronneberger, P.Fischer, and T.~Brox, ``{U-Net}: Convolutional networks for
  biomedical image segmentation,'' in \emph{Medical Image Computing and
  Computer-Assisted Intervention (MICCAI)}, ser. LNCS, vol. 9351.\hskip 1em
  plus 0.5em minus 0.4em\relax Springer, 2015, pp. 234--241.

\bibitem{He_2016_CVPR}
K.~He, X.~Zhang, S.~Ren, and J.~Sun, ``Deep residual learning for image
  recognition,'' in \emph{The IEEE Conference on Computer Vision and Pattern
  Recognition (CVPR)}, June 2016.

\bibitem{7859320}
T.~N. {Sainath}, R.~J. {Weiss}, K.~W. {Wilson}, B.~{Li}, A.~{Narayanan},
  E.~{Variani}, M.~{Bacchiani}, I.~{Shafran}, A.~{Senior}, K.~{Chin},
  A.~{Misra}, and C.~{Kim}, ``Multichannel signal processing with deep neural
  networks for automatic speech recognition,'' \emph{IEEE/ACM Transactions on
  Audio, Speech, and Language Processing}, vol.~25, no.~5, pp. 965--979, 2017.

\bibitem{fischl_2004}
\BIBentryALTinterwordspacing
B.~Fischl, D.~H. Salat, A.~J. van~der Kouwe, N.~Makris, F.~Ségonne, B.~T.
  Quinn, and A.~M. Dale, ``\BIBforeignlanguage{en}{Sequence-independent
  segmentation of magnetic resonance images},''
  \emph{\BIBforeignlanguage{en}{NeuroImage}}, vol.~23, pp. S69--S84, Jan. 2004.
  [Online]. Available:
  \url{http://linkinghub.elsevier.com/retrieve/pii/S1053811904003817}
\BIBentrySTDinterwordspacing

\bibitem{Brainstorm2011}
F.~Tadel, S.~Baillet, J.~C. Mosher, D.~Pantazis, and R.~M. Leahy, ``Brainstorm:
  a user-friendly application for {MEG/EEG} analysis,'' \emph{Computational
  intelligence and neuroscience}, 2011.

\bibitem{Huang1999}
H.~M. X., M.~J. C., and R.~M. Leahy, ``A sensor-weighted overlapping-sphere
  head model and exhaustive head model comparison for {MEG},'' \emph{Physics in
  medicine and biology}, vol.~44, p. 423–440, 1999.

\bibitem{sekihara2001}
K.~{Sekihara}, S.~S. {Nagarajan}, D.~{Poeppel}, A.~{Marantz}, and
  Y.~{Miyashita}, ``Reconstructing spatio-temporal activities of neural sources
  using an meg vector beamformer technique,'' \emph{IEEE Transactions on
  Biomedical Engineering}, vol.~48, no.~7, pp. 760--771, 2001.

\bibitem{tensorflow2015-whitepaper}
\BIBentryALTinterwordspacing
M.~A. et~al., ``{TensorFlow}: Large-scale machine learning on heterogeneous
  systems,'' 2015, software available from tensorflow.org. [Online]. Available:
  \url{http://tensorflow.org/}
\BIBentrySTDinterwordspacing

\bibitem{dinh2019contextual}
C.~Dinh, J.~G. Samuelsson, A.~Hunold, M.~S. Hämäläinen, and S.~Khan,
  ``Contextual minimum-norm estimates (cmne): A deep learning method for source
  estimation in neuronal networks,'' 2019.

\end{thebibliography}

% biography section
% 
% If you have an EPS/PDF photo (graphicx package needed) extra braces are
% needed around the contents of the optional argument to biography to prevent
% the LaTeX parser from getting confused when it sees the complicated
% \includegraphics command within an optional argument. (You could create
% your own custom macro containing the \includegraphics command to make things
% simpler here.)
%\begin{IEEEbiography}[{\includegraphics[width=1in,height=1.25in,clip,keepaspectratio]{mshell}}]{Michael Shell}
% or if you just want to reserve a space for a photo:

% if you will not have a photo at all:

% insert where needed to balance the two columns on the last page with
% biographies
%\newpage

% You can push biographies down or up by placing
% a \vfill before or after them. The appropriate
% use of \vfill depends on what kind of text is
% on the last page and whether or not the columns
% are being equalized.

%\vfill

% Can be used to pull up biographies so that the bottom of the last one
% is flush with the other column.
%\enlargethispage{-5in}

% that's all folks
\end{document}